\documentclass[fleqn,twoside]{article}
\usepackage{amsmath}
\usepackage{graphicx}
\usepackage{epsfig}
\usepackage[usenames,dvips]{color}
\usepackage{cite}
\usepackage{espcrc2}

\usepackage[figuresright]{rotating}
\def\be#1\ee{\begin{equation}#1\end{equation}}
\def\bea#1\eea{\begin{eqnarray}#1\end{eqnarray}}
\def\ba#1\ea{\begin{align}#1\end{align}}
\def\btab#1\etab{\begin{table}[h]\begin{center}#1\end{center}\end{table}}
\def\bfig#1\efig{\begin{figure}[h]\begin{center}#1\end{center}\end{figure}}
\def\bi#1\ei{\begin{itemize}#1\end{itemize}}
\def\bc#1\ec{\begin{center}#1\end{center}}

\newcommand{\ie}{i\varepsilon}

\def \ie{i\varepsilon}

\def\diana{{\sc Diana}}
\def\form{{\sc Form}}
\def\fortran{{\sc Fortran}}

\def\make{{\sc Make}}
\def\kitform{{\sc kitForm3}}
\def\kitfortran{{\sc kitFortran}}

\def\ff{{\sc FF}}
\def\LT{{\sc LoopTools}}

\def\FC{{\sc FormCalc}}
\def\qgraf{{\sc Qgraf}}

\def\aitalc{{\sc \textit{a}{\r{\i}}\raisebox{-0.14em}{T}alc}}

\title{
Automated calculations for massive fermion production with \aitalc
\thanks{Work supported in part by European's 5-th Framework under
  contract HPRN--CT--2000--00149 Physics at Colliders and by the
  Deutsche Forschungsgemeinschaft under contract SFB/TR 9--03.}
}

\author{A. Lorca\address[zeuthen]{Deutsches Elektronen-Synchrotron,
    DESY, Platanenallee 6, 15738 Zeuthen, Germany} and
T. Riemann\addressmark[zeuthen]
}

\begin{document}

\begin{abstract}
The package \aitalc{} has been developed for the automated calculation of
radiative
corrections to two-fermion production at $e^+e^-$ colliders.
The package uses \diana, {\qgraf}, \form, \fortran,  \ff,  \LT, and further
unix/linux tools.
Numerical results are presented for $e^+e^- \to e^+e^-, \mu^+\mu^-, b{\bar s}, t {\bar c}$.
\end{abstract}

\maketitle

\section{\label{sec-intro}INTRODUCTION}
Two fermion production, e.g.
\be
e^+ e^- \rightarrow e^+ e^-,\mu^+\mu^-,\tau^+\tau^-,b{\bar b}, t{\bar
  t},b{\bar s}, t{\bar c},
\label{twofermion}
\ee
is among the reactions to be observed at a future linear $e^+e^-$
collider (LC)
\cite{Aguilar-Saavedra:2001rg}.  
We developed packages for their calculation in the electroweak
Standard Model (SM) (and extensions of it).
The resulting \fortran{} codes may serve as etalons for other, numerical
programs, may be used as electroweak library for some Monte Carlo
program as well as directly used for studying the corresponding
scattering process.

Earlier studies in this connection are
\cite{Fleischer:2003kk,Hahn:2003ab,Gluza:2003nn,Biernacik:2003xv}.
Here, we report on \aitalc{}, a package with a high degree of
automatization for this kind of calculations, and will focus on two
applications: Bhabha scattering and flavour violating fermion pair production.

\section{\label{sec-auto} AUTOMATED CALCULATIONS WITH \protect \aitalc}

\begin{figure*}[hbt]
\begin{center}
\includegraphics[width=416pt]{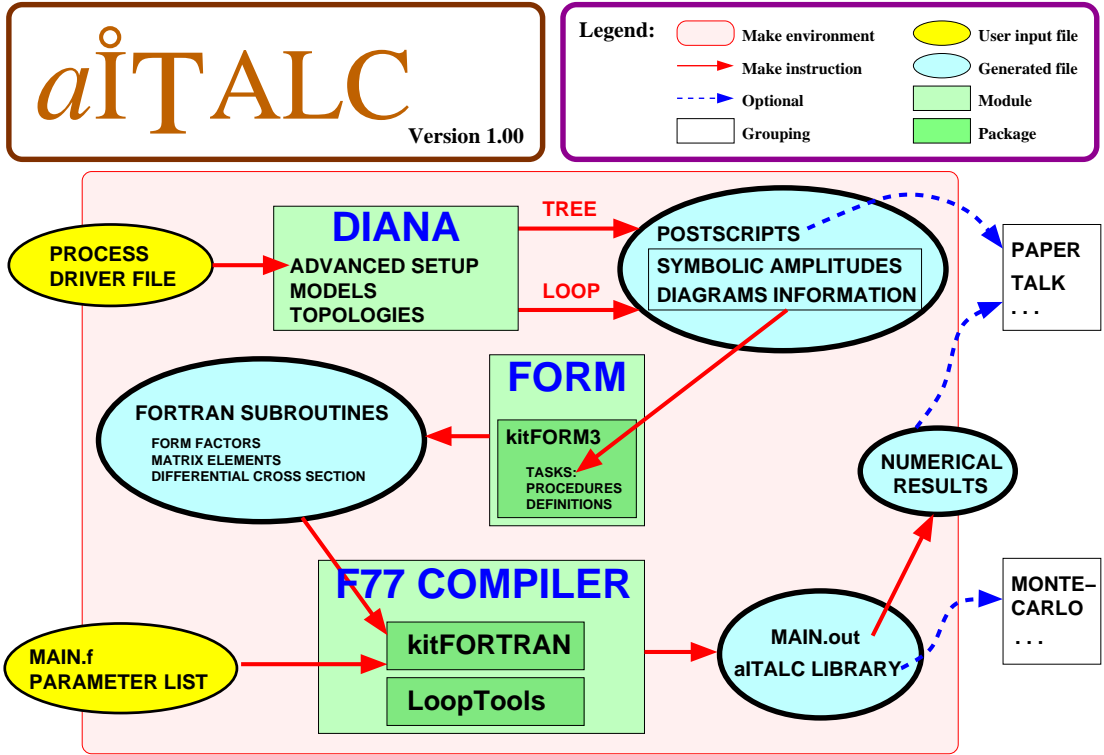}
\vspace{-7pt}
\caption{Flow chart of the \protect \aitalc{} package.}
\label{flowchart}
\end{center}
\end{figure*}

The package \aitalc{} for the automatic calculation of a variety of two
fermion production processes may be obtained from
\cite{Zeuthen-aITALC:2004} where also its installation will be described.  
The logical structure of the package is shown in Figure \ref{flowchart}.
It consists of three modules:
\diana, \kitform{} and \kitfortran.
In order to run and/or modify one of the samples in the 
{\tt example} directory, the user has to choose the process in terms of
incoming and outgoing fermions and the model
lagrangian;
we support two models: {\tt QED.model}
and {\tt EWSM.model}, both with counterterms. 
This is done by modifying the driver file {\tt
  process.ini} in the {\tt sample} directory tree.
Sample processes are: Bhabha scattering ({\tt eeee}), $\mu$ pairs
({\tt muon\_production}), $b$ pairs ({\tt eebb}), $b\bar{s}$ ({\tt
  leLe-bS}), $t\bar{c}$ ({\tt leLe-tC}), etc. Selected cases will
also be part of the public distribution.

The electroweak corrections are organized following
\cite{Denner:1993kt,Bohm:1986rj1}, and
we  keep  all the fermion masses, including $m_e$, by default. 

The user runs the package by {\make} in e.g. {\tt eebb\_user} and
produces (among others) the subdirectories  {\tt tree}, {\tt loop} and  {\tt fortran}.
The {\diana} module uses {\qgraf} v.2.0 \cite{Nogueira:1993ex} 
and {\diana} v.2.35 \cite{Tentyukov:2002ig} and creates symbolic
\form{}-readable output for each of the Feynman diagrams in {\tt
  eebb.in}. 
Moreover, graphical
representations are produced, both stored as encapsulated
postscript figures in {\tt EPS} (single Feynman diagrams) and as postscripts
in {\tt eebbInfo.ps} (with detailed informations for each single diagram)
and in {\tt eebb.ps} (an overview of all contributions).

Then module \kitform{} performs some algebraic simplifications and
determines the matrix elements and form factors. 
This operation takes place separately 
for the tree and loop levels.
The module {\kitfortran}{}
returns the \fortran{} executable file {\tt main.out} and the library
{\tt libaitalc\_v1.a}.
Two \fortran{} files, {\tt parameterlist.hf}\footnote{This file contains only parameter
  declarations to be read in the initialization routine.} and {\tt main.f}, provide the user access to
the input parameters in the model and to the design of numerical
output of the code (i.e. number of data points in the angular
distribution, integrated cross section, running flags, etc.).
By executing {\tt main.out}, the user produces a sample output 
{\tt main.log} with differential and integrated cross sections, as well as
forward-backward asymmetries.
For Bhabha scattering at $\sqrt{s}= 500 $ GeV, a shortened sample is
reproduced in Table \ref{bhabhatable}.

Advanced features are not extremely user friendly, but are still under
development. 
Soft photonic corrections may be included (or not), 
and with {\tt lidentCKM=.true.} quark flavour mixing is discarded with
a diagonal CKM matrix.
One may also
perform a full one-loop tensor integral reduction to the master integrals  
$A_0$, $B_0$, $C_0$ and $D_0$ in the Passarino-Veltman scheme
\cite{Passarino:1979jh}. 
All these possibilities will be described in more detail in a tutorial
to be published soon; see also \cite{Lorca:2003?1}.

\section{\label{sec-results}SELECTED APPLICATIONS}
\subsection{\label{ssec-results1}Bhabha scattering}
Bhabha scattering,
\be
e^+ e^- \rightarrow e^+ e^- ,
\label{bha}
\ee
was one of the first processes to be calculated in
QED \cite{Bhabha:1936xx}, and the 
one-loop corrections in the SM were treated in
\cite{Consoli:1979xw,Bohm:1984yt,Tobimatsu:1986pp,Bardin:1991xe,%
Beenakker:1991mb,Montagna:1993py,Beenakker:1998fi}.  

We are calculating the virtual corrections to massive, low angle Bhabha
scattering at a Linear Collider \cite{Aguilar-Saavedra:2001rg}.
In order to guarantee a precision of $10^{-4}$ of the differential
cross section we need the one-loop corrections in the SM, and the
two-loop corrections in pure, massive QED \cite{Czakon:2004nn}; the 
latter comprise also the squared one-loop corrections
\cite{Fleischer:2002ih}. 

The one-loop SM corrections are determined with \aitalc. 
A shortened example of the \aitalc{} output is given in Table
\ref{bhabhatable}. 
A comparison of our calculation with another one based on {\FC}{}
\cite{formcalc} is reproduced in \cite{Lorca:2003?1}, 
where an agreement is obtained of more than 12 significant digits.

\begin{table*}[hbt]
\caption{Sample output of {\protect \aitalc} for Bhabha
 scattering at $\sqrt{s} = 500$ GeV: Born and 1-loop corrected
  differential cross sections; input data as in
 \cite{Hahn:2003ab,Lorca:2003?1}
}
\begin{verbatim}
# ==================================================================
# aITALC: Version 0.7 by A.Lorca -- T.Riemann
# ==================================================================
#cos(theta) dcs(BORN)               ...  dcs(BORN+Q+W+soft)      ...
-.90000     0.2169988288109205E+00  ...  0.1934450785268578E+00  ...
-.50000     0.2613604305853236E+00  ...  0.2387066977233451E+00  ...
0.00000     0.5981423072503301E+00  ...  0.5466771794694227E+00  ...
0.50000     0.4212729493916255E+01  ...  0.3813007881789546E+01  ...
0.90000     0.1891603223322704E+03  ...  0.1729283490665079E+03  ...
# ------------------------------------------------------------------
\end{verbatim}
\label{bhabhatable}
\vspace*{-08mm}
\end{table*}

\subsection{\label{ssec-results2}At the $Z$-peak}
\begin{figure}[htb]
\includegraphics[scale=.76]{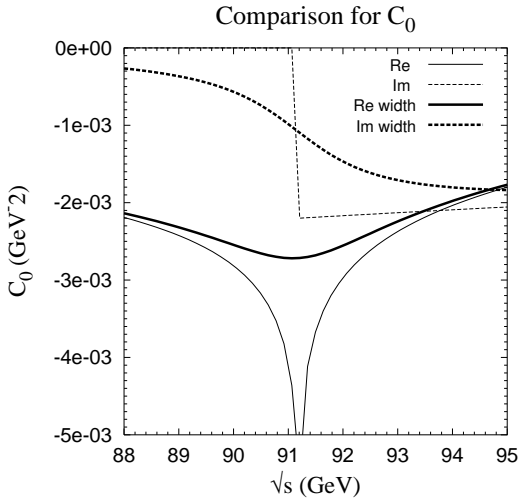} 
\vspace{-9mm}
\caption{%
The 
infrared finite function
$C_0(s,m_{\mu}^2,m_{\mu}^2;m_Z^2,0,m_{\mu}^2)$ as a function 
of $\sqrt{s}$.}
\label{figboth}
\vspace{-6mm}
\end{figure}

For the numerical evaluation of the one-loop functions we use
{\tt LoopTools} v.2.0 and follow the conventions given in the manual
\cite{Hahn:1998yk2,vanOldenborgh:1991yc,Hahn:2001??}.
{\tt LoopTools} relies on the package {\ff}{}
\cite{vanOldenborgh:1991yc}. 
In contrast to \ff, it does not support the case of complex masses.  
Near the $Z$-peak, the Breit-Wigner $Z$ propagator has to be used,
\be 
\frac{1}{s-m_Z^2+\ie} \rightarrow
\frac{1}{s-m_Z^2+i\Gamma_Z m_Z},
\label{zpropagator}
\ee
with a width parameter $\Gamma_Z$.
Flag {\tt lwidth} activates the following changes in \aitalc:
\begin{itemize}
\item Apply substitution (\ref{zpropagator}) in Feynman rules;
\item Avoid double counting by discarding the $Z$ self-energy diagrams
  resummed in $\Gamma_Z$;
\item Use loop integrals with complex $Z$ mass.
\end{itemize}
In general we use for the numerics Passarino-Veltman functions of
scalar, vector, and tensor type.
In the neighbourhood of the $Z$ resonance, though,
 we perform the tensor reduction for
the $\gamma-Z$ dependent functions $D_{ij}(...)$, and have
to calculate the following functions with complex $Z$ mass: 
$B_0$, $C_0$ and $D_0$.
The $A_0$ and $B_0$ are trivial, and the infrared divergent $D_0$
was calculated in \cite{Beenakker:1990jr}.
The tensor reduction of the 
tensor integrals related
to the $\gamma-Z$ box diagrams leads, by shrinking of internal lines, to
infrared safe two- and three-point functions with complex $Z$ mass. 
They are generally treated in \cite{'tHooft:1979xw} and available in the
 \ff{} package
\cite{vanOldenborgh:1991yc}%
\footnote{It has to be noted that the
packages \ff{} and \LT{} v.2.0 \cite{Hahn:1998yk2} are not compatible
and the user has to either drop one of them or modify the internal
source code.  
Therefore, presently \aitalc{} renames part of the \LT{} subroutines.
After this workshop, {\tt LoopTools} v.2.1 (29 June 2004) was
  released and is now prepared for the treatment of complex masses.}. 
We performed an independent calculation of the only nontrivial one,
$C_0(\ldots;m_Z^2-i \Gamma_Z m_Z,0,m_f^2)$, and got perfect agreement
with \ff. 
For an implementation and comparison between the non-width
and width cases see Figure \ref{figboth}.

\begin{figure}[htb]
\includegraphics[scale=.75]{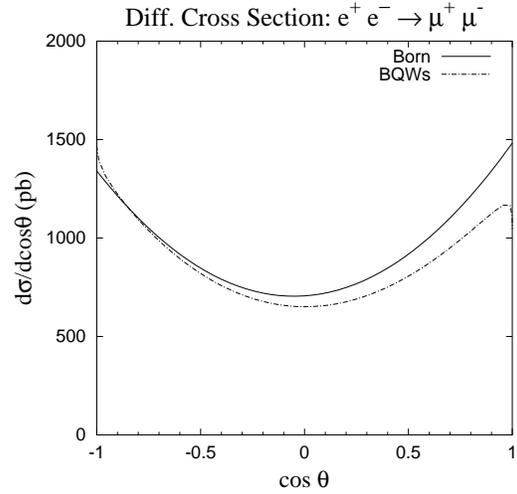}
\vspace{-9mm}
\caption{%
Muon production at the $Z$-peak.
}
\label{dcsmuoncMMZ}
\vspace{-2mm}
\end{figure}

As an example, the  differential cross section for 
\mbox{$e^+e^- \to \mu^+\mu^-$} around  the $Z$-peak is presented in Figure
\ref{dcsmuoncMMZ}.

\subsection{\label{ssec-results3}Flavour number violation}
\begin{figure}[hbt]
\includegraphics[scale=0.75]{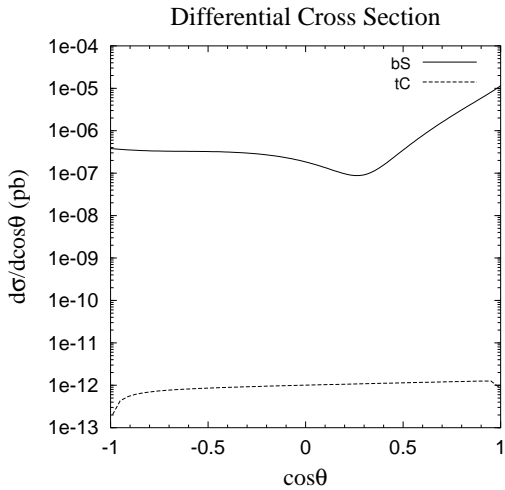}
\vspace{-9mm}
\caption{%
Differential cross sections for
  $e^+e^-\rightarrow b \bar{s}$ and $e^+e^-\rightarrow t \bar{c}$ at $\sqrt{s} = 200$ GeV; input data as
  in \cite{Huang:1999yu}.
}
\label{dceebs}
\vspace{-6mm}
\end{figure}

\begin{figure}[hbt]
\includegraphics[scale=0.75]{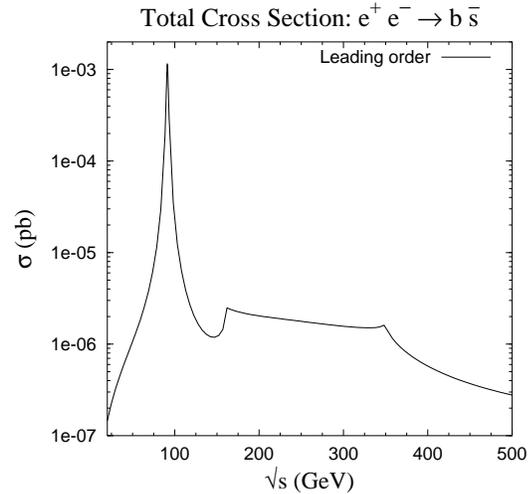}
\vspace{-9mm}
\caption{%
Integrated cross section for 
  $e^+e^-\rightarrow b \bar{s}$ as a function of $\sqrt{s}$; input data as
  in \cite{Huang:1999bt}.
}
\label{ics_bs}
\vspace{-4mm}
\end{figure}

Topics of particular interest are  flavour changing neutral current processes
at $e^+ e^-$ colliders since they are forbidden in the Born
approximation of the Standard Model and might indicate New Physics.
At one-loop order, they may occur in the SM if the fermions of
different flavour have different masses and are mixing.
Usually one calculates simply the flavour changing $Z$ decay rate because the
chances to observe an effect are largest at the $Z$-peak.
The predictions in the minimally extended  Standard Model are tiny
(see e.g. \cite{Mann:1984dv,Ganapathi:1983xy,Illana:2000ic}),
but may be much larger with supersymmetry (see e.g. \cite{Illana:2002tg}).
Recently, there were also studies of the complete scattering processes
\cite{Huang:1999bt,Huang:1999yu}: 
\be
e^+ e^- \to  t \bar{c}, \quad b \bar{s}.
\label{fermionv}
\ee
The mass effects, but also both $Z$ and $\gamma$ exchange, and the
$WW$ box are taken into account.
We recalculated the corresponding rates with \aitalc.  
Figure \ref{dceebs}  shows our 
differential cross sections for (\ref{fermionv}).
By inspection of the corresponding plot in
\cite{Huang:1999yu}, one realizes a normalization difference of about $3/2$
or  $\pi/2$ by which our cross section is larger.
Using the input data of \cite{Huang:1999yu}, we get:
\begin{eqnarray}\nonumber
\sigma_{b{\bar s}}( \sqrt{s}=m_Z ) &=& (1.136 \pm 0.001) \;
\mathrm{f{}b} ,
\\ \nonumber
\sigma_{b{\bar s}}( \sqrt{s}=200\mathrm{GeV} ) &=& (2.033 \pm 0.002)
\; \mathrm{f{}b} .
\end{eqnarray}
For the other channel, we agree within the accuracy of the
figures. 
Figure \ref{ics_bs} shows also the total cross section for $b\bar{s}$.

\section{\label{sec-sum} CONCLUDING REMARKS}
We plan to include into \aitalc{} also processes which have
contributions from Feynman diagrams with five-point functions.
This would allow us to calculate the one-loop corrections to the
radiative Bhabha process, $e^+e^- \to e^+e^-\gamma$.

Further, it would be very important to have a proper treatment of
Majorana particles with {\qgraf} and {\diana}, thus allowing
also the treatment of supersymmetric model files. 
\section*{Acknowledgements}
We thank Thomas Hahn for 
producing a Bhabha \fortran{} code and for
communications when we performed numerical
comparisons with this code.
We also thank J. Fleischer and M. Tentyukov for their cooperation
whenever we had some problems with an effective use of {\diana}.

\end{document}